\documentclass[12pt]{iopart}

\usepackage{graphicx}
\begin{document}

\title[Interactive L\'evy Flight in Interest Space]{Interactive L\'evy Flight in Interest Space}

\author{Fanqi Zeng$^1$, Li Gong$^1$, Jing Liu$^1$, Jiang Zhang$^{1,*}$, Qinghua Chen$^{1,*}$, Ruyue Xin$^1$}

\address{$^1$ School of Systems Science, Beijing Normal University, Beijing 100875, P.R.China}
\address{$^*$ zhangjiang@bnu.edu.cn, qinghuachen@bnu.edu.cn}

\begin{abstract}
Compared to the well-studied topic of human mobility in real geographic space, very few studies focus on human mobility in virtual space, such as interests, knowledge, ideas, and so forth. However, it relates to the issues of management of public opinions, knowledge diffusion, and innovation. In this paper, we assume that the interests of a group of online users can span a Euclidean space which is called interest space, and the transfers of user interests can be modeled as the L\'evy Flight on the interest space. To consider the interaction between users, we assume that the random walkers are not independent but interact each other indirectly via the digital resources in the interest space. The model can successfully reproduce a set of scaling laws for describing the growth of the attention flow networks of real online communities, and the ranges of the exponents of the scaling are similar with the empirical data. Further, we can infer parameters for describing the individual behaviors of the users according to the scaling laws of the empirical attention flow network. Our model can not only provide theoretical understanding on human online behaviors, but also has wide potential applications, such as dissemination and management of public opinions, online recommendation, etc.
\end{abstract}

%
\vspace{2pc}

%
%
%

\section{Introduction}

Everything is moving. To understand the mobility patterns for human kinds is of great importance since it relates to epidemics\cite{Grenfell2001Travelling,Belik2011Natural}, urban planning\cite{Ratti2006Mobile,Lathia2012The}, and other issues in modern city\cite{Wang2009Understanding,Ratti2010Redrawing,Lathia2011Mining,Yuan2012Discovering,Cacciapuoti2012Human,Santi2013Taxi}. Lots of studies on human mobility in real space have been made in past decades\cite{Brockmann2006The,Gonz2009Understanding,Song2010Modelling}. For instance, it is found that L\'evy Flight\cite{Viswanathan1996L,Lomholt2008L,Raposo2009L,Rhee2011On}, one of the most famous random walk model which is significantly distinguished from Brownian motion\cite{Bartumeus2002Optimizing}, can be used to characterize human movements. However, human mobility does not only take place in real space exclusively, but also in virtual space\cite{Lambiotte2008Geographical,Liang2009Avatar}. For example, our consciousness always jumps between different ideas, which can be understood as a virtual movement in interest space\cite{Wu2007Novelty}. A large amount of users surfing on an online community, and jumping between different posts, can be also understood as collective movements in interest space\cite{Wu2013The,Sasahara2013Quantifying,Kenett2014Discovering,Hawelka2014Geo}. Although virtual space seems to be less solid than physical space, the study of it is of great significance because it may help us to understand the issues of psychology therapy\cite{Li2014Game}, dissemination\cite{Etter2013Where} and management of public opinion\cite{Watts2007Influentials}, online recommendation\cite{Smith2005Online}, and so on.

Some important conclusions for collective users in online community have been achieved. For example, conventional studies of human dynamics usually concerned the statistics on a single interest (digital resource) such as access time, frequency, and so forth, or the distributions of the number of visiting pages and the decay patterns of popularity. Besides, human being is a social animal\cite{P2016The}. Thus, a large amount of attention is paid on how people interact, correlate, and connect each other in the studies of social networks\cite{Liben2003The,Mislove2007Measurement,Kumar2010Structure,Cho2011Friendship}. The interaction and correlation between people can also be reflected by statistical laws\cite{Brockmann2006The,Gonz2009Understanding,Song2010Modelling}. For example, the super-linear scaling of productivity is found in both cities and online communities\cite{Lathia2012The, Wu2007Novelty}, which means cooperation between people may promote the growth of the per capita productivity for human organizations in a faster rate than their sizes. While, the widely existed sub-linear scaling law of diversity of places or interests indicates a slower increase of diversification. This is also an emergent indirect results of interactions between people.     

Although human dynamics and complex networks have drawn a wide attention, little concern is made with the sequential movements of users on interests because the concept of the interest space is not apparent. Recently, some attempts have been made to visualize the virtual space by attention flow network model\cite{Shi2015A,Zeng2017Visualizing}, on which nodes represent digital resources (posts, tags, and articles etc.). The network is constructed according to the collective behaviors of a large number of users. However, the attention flow network is built according to the data of users\cite{Wu2013The, Frank2013Happiness}, i.e., the representation of the interest space rather than the space itself.  

In this paper, we focus on collective movements of users in their interest space. Here, we assume that all the possible interests of users span a Euclidean space, in which adjacent points standing for similar interests, and users performing random walks of L\'evy Flights in the space. However, naive L\'evy Flight model\cite{Viswanathan1996L,Raposo2009L} can not reproduce the required scaling laws because of the absence of interactions. Thus, we build an interactive L\'evy Flight model, in which, interactions occur between users indirectly bridged by digital resources. The model successfully reproduces all the concerned scaling laws, and the range of scaling exponents can be also calibrated by adjusting the parameters in the model. Our model can not only deepen our understanding, but also may largely improve the accuracy of predictions on user behaviors\cite{Liang2009Avatar,Yuan2015How}, leading to wide applications on recommendation\cite{Smith2005Online,Pennacchiotti2011Investigating}, searching\cite{Cole2015User}, user profiling\cite{Middleton2004Ontological}, etc.

\section{Model}

\begin{figure}[ht]
	\centering
	\includegraphics[width=\linewidth]{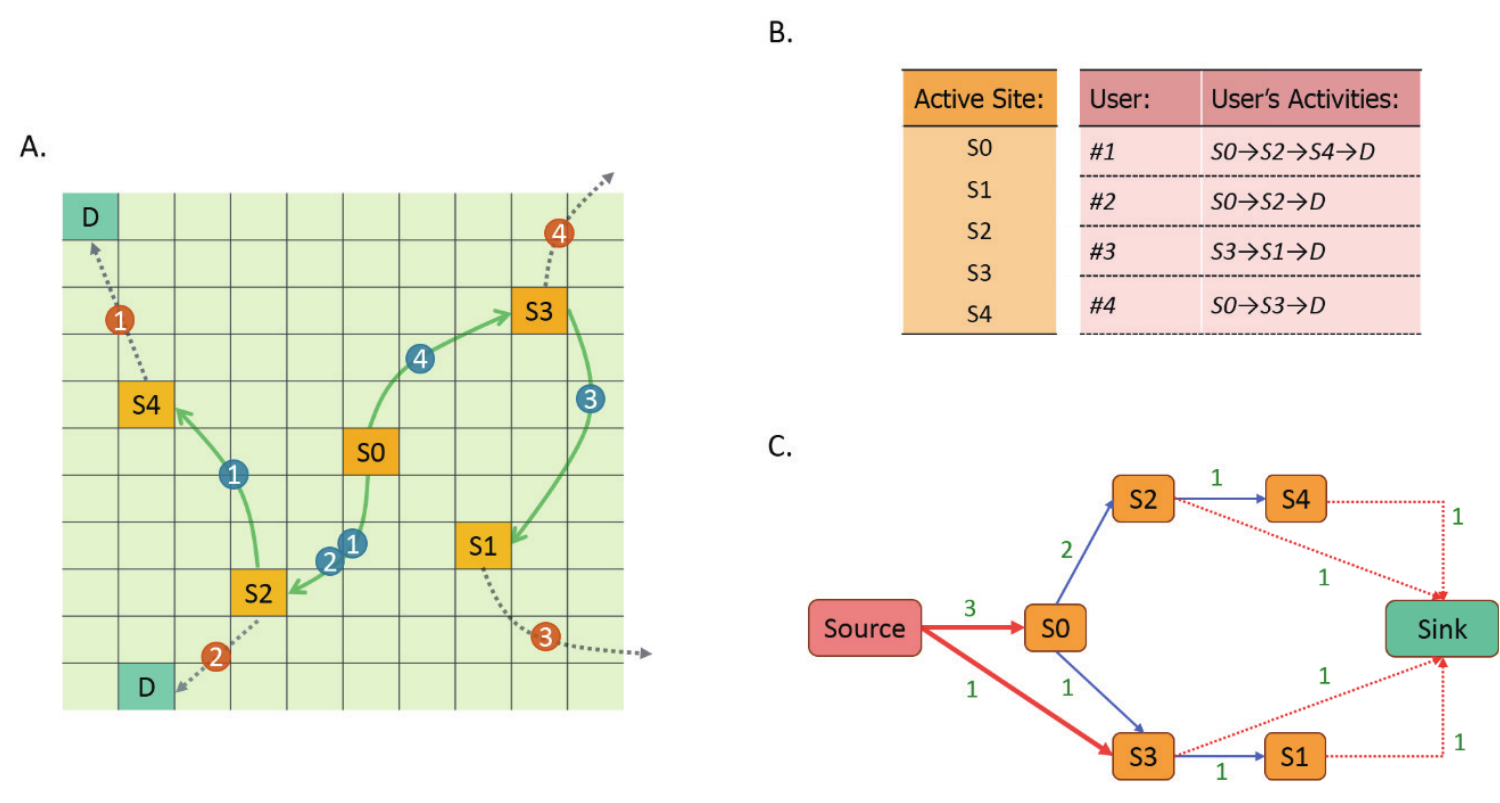}
	\caption{\textbf{The Interactive L\'evy Flights model in a 2-dimensional space.} (A) the active sites and the trajectories. (B) generated by the random walkers in (A), and the constructed attention flow network. (C) according to the trajectories in (B). In (A), $S0$, $S1$,..., $S4$ are active sites with digital resources; Circled numbers represent users who implementing L\'evy flights. (D) represents the places where the users jumping to but no digital resources existing on, and the users quitting consequently.}
	\label{fig:levy_model}
\end{figure}

To understand the scaling phenomena of production and diversity, and the relationship between users and digital resources, we construct an interactive L\'evy Flight model to simulate user behaviors. Let’s consider an online community (such as Baidu Tieba, Stack Exchange or Flickr, etc.), which containing a large number of registered users. All their interests can span an interest space which is modeled by a 2-dimensional Euclidean plane (see figure \ref{fig:levy_model}-(A)). In which, each cell represents one possible interest, such as a kind of music style, or a type of article, etc. Two cells are adjacent standing for they representing similar contents. Meanwhile, articles, tags, Q\&As, and so forth digital resources generated by users can be projected onto this interest space. We use $C(X,t)$ to denote the number of units projected at $X$ and time $t$, where $X$ is the coordinate of the cell. The cell occupied by at least one unit of digital resource, i.e., $C(X,t)>0$, is called an Active Site meaning that the resource has the corresponding theme as the interest. In figure \ref{fig:levy_model}-(A), $S0$, $S1$,...,$S4$ represent active sites and occupied by digital resources. When a digital resource is generated by a user, it will be projected on the space and be able to be visited or read by other users. Thus, all users interact each other indirectly through the active sites.

Users’ sequential behaviors such as browsing, posting, Q\&A, and others in a session can be viewed as a walk in the interest space. We assume that the user's walk satisfies the 2-dimensional L\'evy Flight law, meaning that the movement pattern of a user is basically random and the probability density distribution function of the movement distance $l$ in one jump is a power law:

\begin{equation}
P(l) \sim \l^{-\lambda}\label{eq1}
\end{equation}

This movement rule describes how a user’s interest transfers: user's interest will stay in a narrow area for a long time, but occasionally implement a long-range jump with a small probability. $\lambda$ (values are in [1,3]) is the exponent of the L\'evy flight, it characterizes how wide are the interests of users. If $\lambda$ is small, then the users may frequently perform long range random jump, representing they have wide interests, and vice versa. Thus, $\lambda$ is a parameter of a trade-off between familiarity and novelty. Users usually consume some familiar topics, but they also require some new information occasionally to visit. In figure \ref{fig:levy_model}-(A), the arcs labeled with same numbers represent the flights of one user. For example, the user with label $1$ travels $S0$, $S1$, and $S4$ sequentially.

Next, we consider the interaction between users. We know that if a community already has abundant digital resources (such as many posts in a forum), users will continue to visit these resources. Otherwise they will lose their interests and quit quickly. In order to characterize this feature, without loss of generality, we assume that the user can jumps continually and randomly from a cell $X$ as long as $X$ is active (there are at least one unit of digital resource), otherwise, he(she) will go out of the space from $X$. We denote the position of user $i$ at time $t$ is $X_t$ and it is $\emptyset$ when the user quit the community, so we have

\begin{equation}
\label{cases}
X_{t}^{i}=\cases{X_{t-1}^{i}+\xi&if $C(X_{t-1}^{i}, t-1) > 0$\\
				 \emptyset&if $C(X_{t-1}^{i}, t-1) = 0$\\}\label{eq2}
\end{equation}

where, $\xi$ is a random number following equation (\ref{eq1}).
On the other hand, each user will generate new digital resource with a certain probability in the process of random walk. That is, if the user $i$ jumps to the cell $X$, he(she) will add a new digital resource at $X$ with a certain probability $p$. Thus, we have:

\begin{equation}
C(X_{t}^i, t) = C(X_{t-1}^i, t-1)+\eta\label{eq3}
\end{equation}

where, $\delta$ is the Dirac delta function, it equals to $1$ only if its component is $0$; and $\eta$ is a random number following $0-1$ distribution with a probability $p$ to be $1$, and $N$ is the total number of users who performs L\'evy Flights in the space.

Next, we consider the situation with $N$ users. Suppose in one simulation (a session), $N$ users are set on the origin of the interest space, and they begin to implement L\'evy Flights from the origin simultaneously. Although they don’t interact directly, they can interfere each other via the active sites. The simulation ends at time $T(N)$ when all users exit the space. Apparently, $T(N)$ will increase with $N$ implying that the indirect interactions can keep users living in the community for longer time since the probability that they encounter each other increases. One trajectory can be generated for one user in his(her) lifespan in the community. 

To observe the collective behaviors of these L\'evy Flighters, we construct an attention flow network as shown in figure \ref{fig:levy_model}-(B). The so called attention flow network is an open flow network\cite{Guo2015Flow} in which nodes representing digital resources (active sites) and weighted directed links representing transition flows between two nodes formed by the collective behaviors of the random walkers. In the model, the weight of the edge connecting active sites $X$ and $Y$ can be defined as:

\begin{equation}
W(X, Y)=\sum_{\tau=1}^{T-1}\sum_{i=1}^{N}\delta(X_{\tau}^{i}-X)\cdot\delta(X_{\tau+1}^{i}-Y)\label{eq4}
\end{equation}

Two special nodes are added to represent the environment which are the source and the sink. When a random walker starts to jump to an active site $X$ in the interest space, a unit of flux from the source to the node $X$ is added to the attention flow network. On the other hand, a unit of flux from node $X$ to the sink is added if the last site of a random walker’s visit is $X$. The attention flow network can characterize the collective properties of a large number of users for both the simulated model and the empirical data.

\section{Simulation}
To validate our model, we study how the network properties will change as the size of the system increases, and to see if the same scaling laws can be reproduced by our model. Here, we use the total number of users $N$ as the measure of the size. In fact, this quantity is also the total influx to the attention network for a given simulation. Following that, we will focus on how the macroscopic properties change with $N$. 

\begin{figure}
	\centering
	\includegraphics[width=0.9\linewidth]{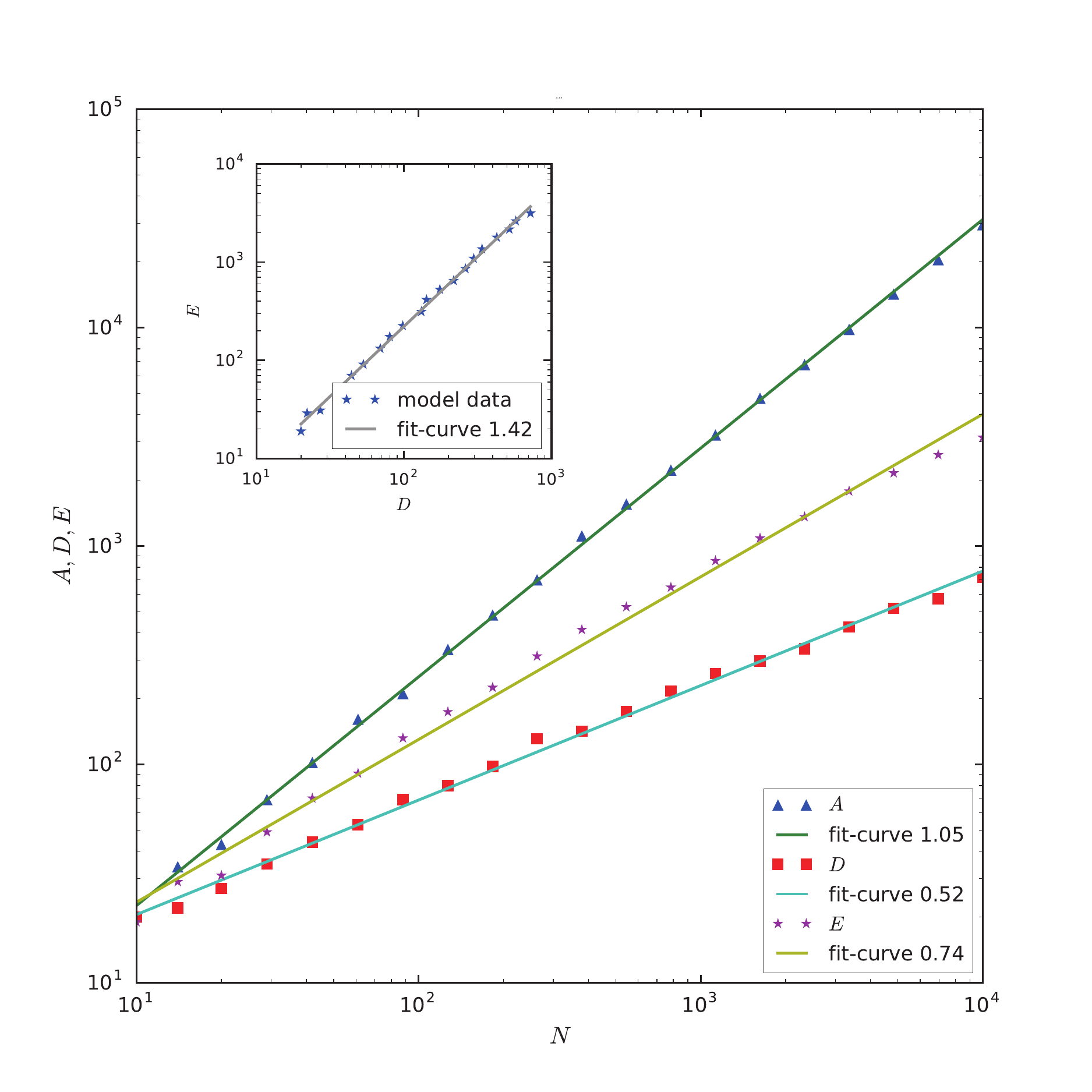}
	\caption{\textbf{The scaling of the simulated attention flow network constructed by interactive L\'evy flight model.} Three quantities $A,D,E$ scale with $N$ with different exponents. Inset shows the densification law, i.e., the scaling of $E$(the number of edges) with $D$(the number of nodes) of the attention flow network.}
	\label{fig:levy_coeffecients}
\end{figure}

Here, we focus on three basic macroscopic variables. First,
\begin{equation}
A=\sum_{X}\sum_{Y}W(X, Y)
\end{equation}
measures the activity of the community, it is defined as the total number of transitions of interests (jumps). Second,
\begin{equation}
D=\sum_{X}\sum_{t=1}^{T}\delta(C(X, t))
\end{equation}
is the total number of active sites, or the total number of nodes in the network. It measures the diversity of interests for all members in the community. Third,
\begin{equation}
E=\sum_{X}\sum_{Y}\delta(W(X, Y))
\end{equation} 
is the total number of edges of the network, and it measures the diversification of interests transitions. According to our simulated data, all three variables scale to 
$N$ with different exponents, i.e.,
\begin{eqnarray}
A \sim N^{\alpha}\\
D \sim N^{\beta}\\
E \sim N^{\gamma}
\end{eqnarray}
Where $\alpha$, $\beta$, and $\gamma$ are exponents characterizing the relative growth speed of the quantities to the size of the system as shown in figure (\ref{fig:levy_coeffecients}).

\begin{figure}
	\centering
	\includegraphics[width=\linewidth]{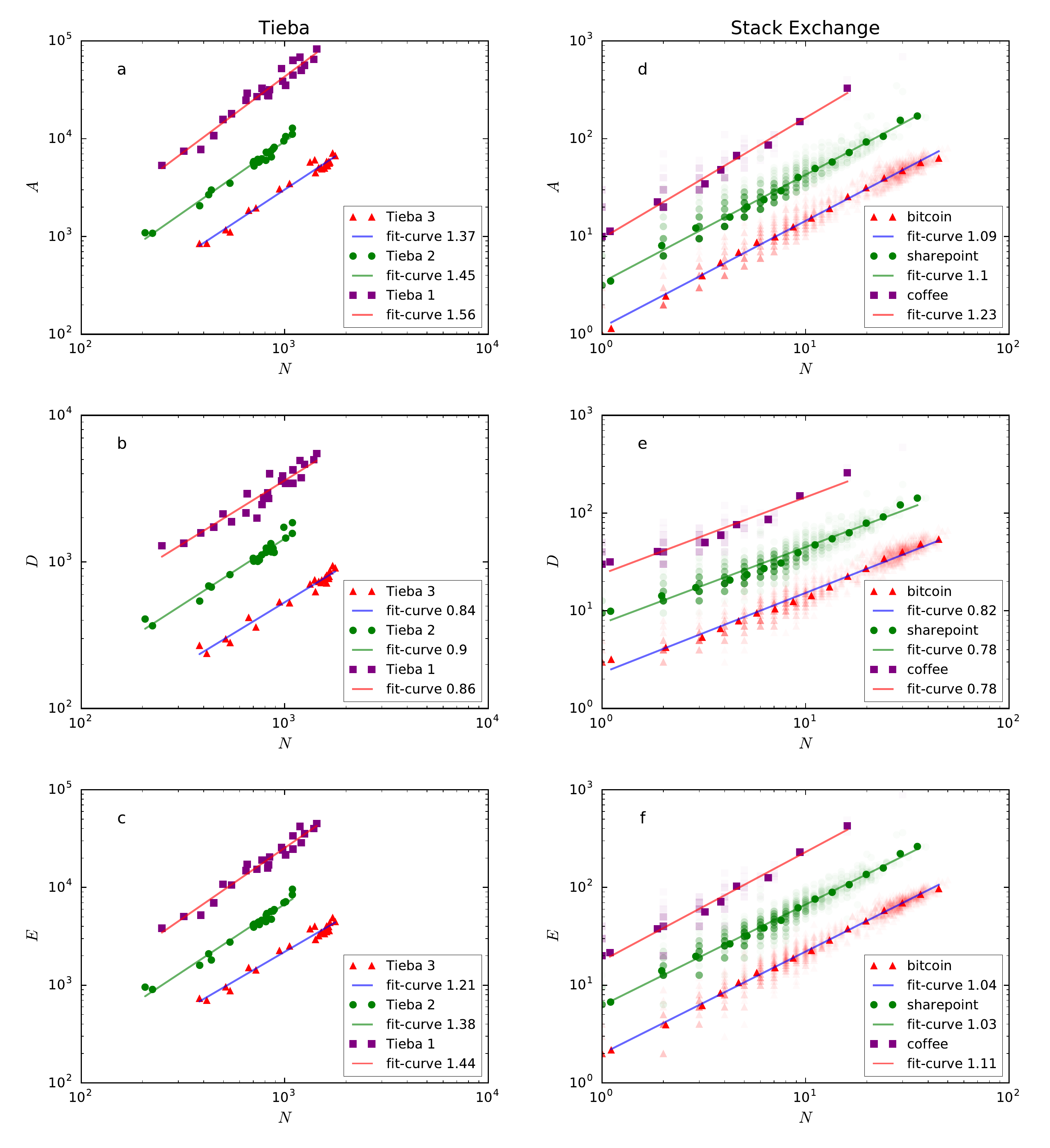}
	\caption{\textbf{The scaling of empirical attention flow networks constructed by the behaviors of users in representative online communities.} We plot $A(a,d)$, $D(b,e)$, and $E(c,f)$ versus $N$ for both Baidu Tieba (a,b,c) and Stack Exchange(d,e,f). Three communities are selected from Tieba and Stack Exchange.}
	\label{fig:real_coeffecients}
\end{figure}

To compare with the simulated data, we also plot the empirical scaling laws for the same quantities on three representative online communities, Baidu Tieba (each jump represents a click behavior, see figure \ref{fig:real_coeffecients}-(a,b,c)) and Stack Exchange(each jump represents an answering behavior, see figure \ref{fig:real_coeffecients}-(d,e,f )).

\begin{figure}
	\centering
	\includegraphics[width=\linewidth]{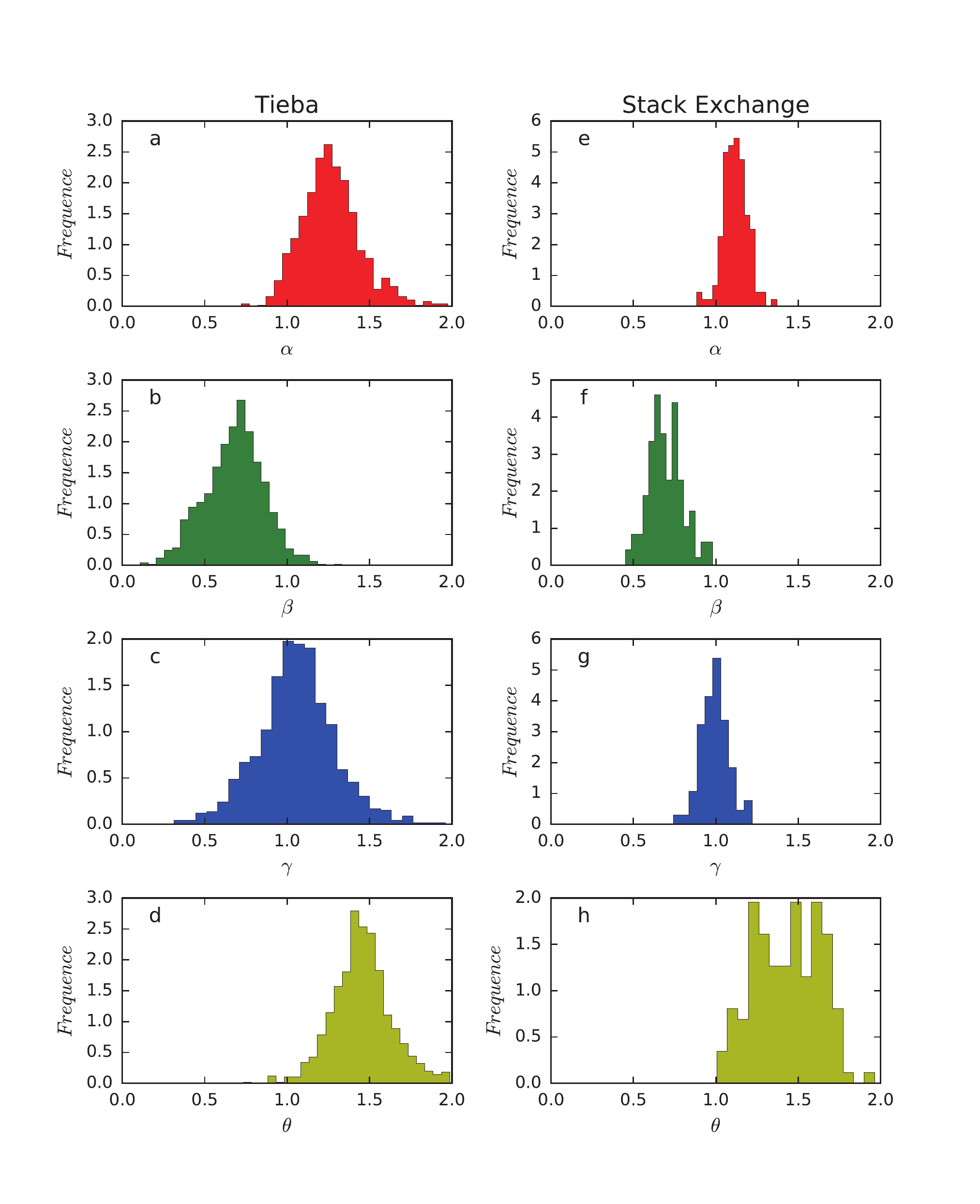}
	\caption{\textbf{The distributions of exponents for 1000 Baidu Tieba (a,b,c,d) and 136 communities on Stack Exchange (d,e,f,g).} Different rows correspond different exponents.}
	\label{fig:real_coeffecients_distribution}
\end{figure}

We observe that all the communities follow the same scaling laws as the simulated results, and the values of exponents are also similar. 
First, we notice that the exponents $\alpha$ are always larger than 1.0 for different $p$ values in simulations. This observation also holds for empirical data. As shown in figure \ref{fig:real_coeffecients_distribution}-(a, b), we systematically calculate the exponents of $1000$ Baidu Tiebas and $136$ communities in Stack Exchange and plot the distribution of exponents.

It is clear that the distribution of the four exponents are nearly normal distribution, in which $\alpha$ is right skew and the average value of $\theta$ is approximately $1.25$ which is larger than $1.0$ significantly. Some small Tiebas’ exponents are less than $1.0$ since their scaling properties are not statistically significant.

We further confirm the super-linear relationship between $A$ and $N$ for more online communities as shown in table \ref{table:average_coeffecients}. All the exponents are larger than $1$. Among which, the communities with intensive interactions between users always have larger exponents like Baidu Tieba, Stack Exchange, and Digg. 

\begin{table}
	\caption{\label{table:average_coeffecients} \textbf{Average exponents $\alpha$ for six representative communities}}
	\begin{tabular*}{\textwidth}{@{}l*{15}{@{\extracolsep{0pt plus 12pt}}l}}
		\br
		Sites     & Baidu Tieba & Stack Exchange &  Delicious & Flickr & Yelp tip & Digg\\
		\mr
		$\alpha$  & 1.27        & 1.12	         & 1.07       & 1.04   & 1.11     & 1.16\\
		\br
	\end{tabular*}
\end{table}

Actually, we can understand the exponent $\alpha$ as an indicator to measure the intensity of the social interactions between users for one community. According to (\ref{eq1}) we derive:
\begin{equation}
\frac{A}{N} \sim N^{\alpha-1}
\end{equation}
That indicates the average number of jumps increases with the size of the system $N$ if $\alpha>1$. And the relative speed increases with $\alpha$. Therefore, if $\alpha$ is big, the average activities generated by the users will be sensitive to the total number of users. This characterizes the nonlinearity of the interactions of users.

\begin{figure}
	\centering
	\includegraphics[width=\linewidth]{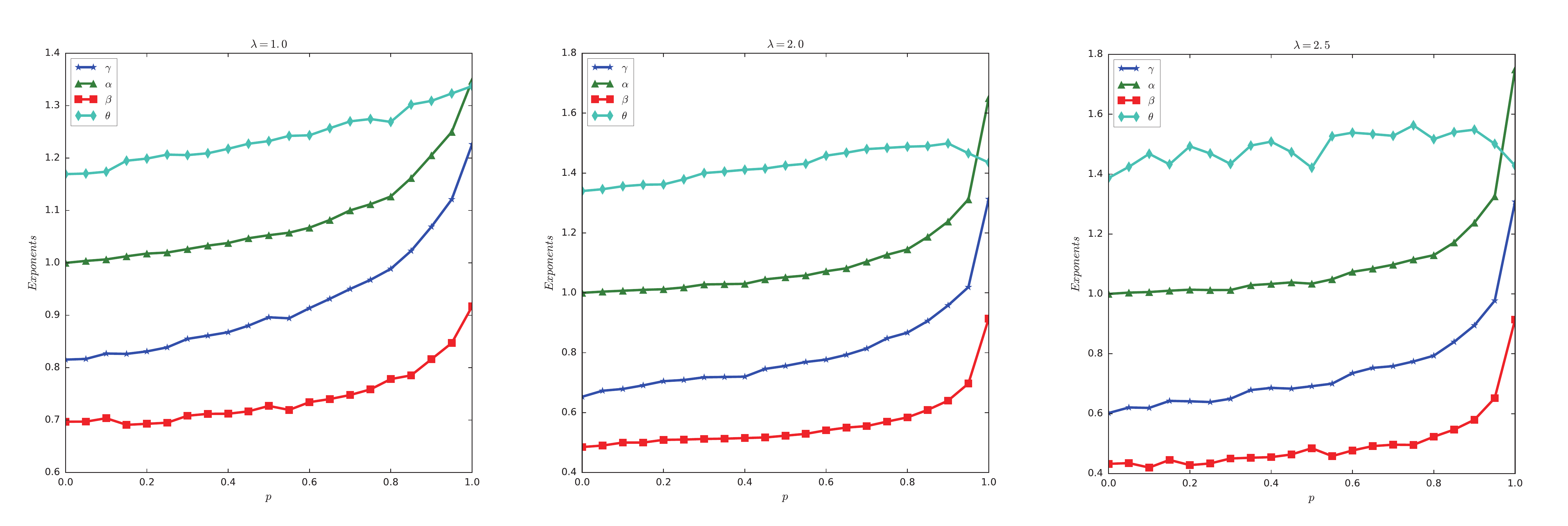}
	\caption{\textbf{The dependency of all exponents on $p$ and how the relations changes with $\lambda$}}
	\label{fig:levy_dependency}
\end{figure}

To compare, we investigate the possible intervals of $\alpha$ for our simulations. As shown in Fig.5, when the exponent $\alpha$ increases as the probability $p$, which means as the propensity that user generating activities increases, the average intensity of interaction also increases. 

If we understand the activity as a kind of production of users, then the exponent $\alpha$ characterizes the productivity of the bunch of users. If it is easy for users to express their interests ($p$ increases), the online community is more productive.

Next, we analyze the exponent of $\beta$, the scaling between the number of nodes of the attention flow network and the size of the system. This scaling law indicates how the diversification of the digital resources generated by the users changes with the size. We found both for simulation(see figure \ref{fig:real_coeffecients}) and empirical data(see figure \ref{fig:levy_dependency}), the exponents are significantly less than one which indicates a sub-linear scaling between diversity and the size of the system. This sub-linear scaling is always observed in other complex systems.

The total number of edges $E$ on the attention flow network measures the diversification of distinct transitions between pairs of nodes. However, there is a large deviation for the exponent $\gamma$. Super linear and sub linear are both possible for different communities. There is a transition from sub-linear to super-linear for simulations.

When $p$ increases, both exponents ($\beta$ and $\gamma$) for diversification increase. That means the propensity that user generating contents can accelerate the relative speeds that diverse contents are produced compared to the size of the system. Thus, the average distinct contents generated increase with the size of the system.
It is interesting to observe another scaling behavior between $E$ and $D$,
\begin{equation}
E \sim D^{\theta}
\end{equation}
a super-linear can be observed. This phenomenon is observed for a large number of networks which is named as densification phenomenon. Our model can successfully reproduce this phenomenon and the exponent $\theta$ fluctuates around $1.5$. All the ranges of exponents for models are consistent with the ones in empirical data, which implies that our model can capture the scaling behaviors in data.

We further test how the exponent of L\'evy Flight influences the other exponents. The results are shown in figure \ref{fig:levy_dependency}. The qualitative characteristics of the dependence between the exponents and $p$ do not change dramatically, however, the range of exponents change is different. 

We also note that the range of the fluctuation of $\lambda$ is relatively small for different $p$, but changes dramatically with different exponent $\lambda$. Thus, we guess that the exponent of $\gamma$ exclusively depends on $\lambda$, and we suppose that this dependence can be used to infer the value of $\lambda$ for a real community.

\section{Inferences for parameters $\lambda$ and $p$}
Next, we will infer the parameters $\lambda$ and $p$ from empirical exponents by using maximum likelihood principle for each community. We suppose the real exponents $\alpha$, $\beta$, and $\theta$ are random sampled from the model. And the exponents follow normal distributions with centers determined by the model and standard deviation $\sigma$ for given $p$ and $\lambda$, that is:

\begin{equation}
P(\alpha, \beta, \theta|p, \lambda) \sim \exp\left(-\frac{(\alpha-\alpha(p, \lambda))^2+(\beta-\beta(p, \lambda))^2+(\theta-\theta(p, \lambda))^2}{\sigma^2}\right)
\end{equation}

where, $\alpha(p,\lambda)$, $\beta(p,\lambda)$, and $\theta(p,\lambda)$ represent the exponents generated the model for given parameter $p$ and $\lambda$ which can be read from the dependency of figure \ref{fig:levy_dependency}. To infer $p$ and $\lambda$ from given empirical measure of $\alpha_i$, $\beta_i$, and $\theta_i$, we attempt to maximize the likelihood probability (eq.\ref{equmax} ), that is:

\begin{equation}\label{equmax}
p,\lambda={\mathop {\arg \max}\limits_{p, \lambda} P(\alpha, \beta, \theta|p, \lambda)}
\end{equation}

So, we need to minimize the distance:
\begin{equation}
D=\sqrt{(\alpha-\alpha(p,\lambda))^2+(\beta-\beta(p,\lambda))^2+(\theta-\theta(p,\lambda))^2}
\end{equation}
That is, we should find the most probable parameters $p$ and $\lambda$ so that the simulated exponents are closest to the empirical ones.

\begin{figure}
	\centering
	\includegraphics[width= \linewidth]{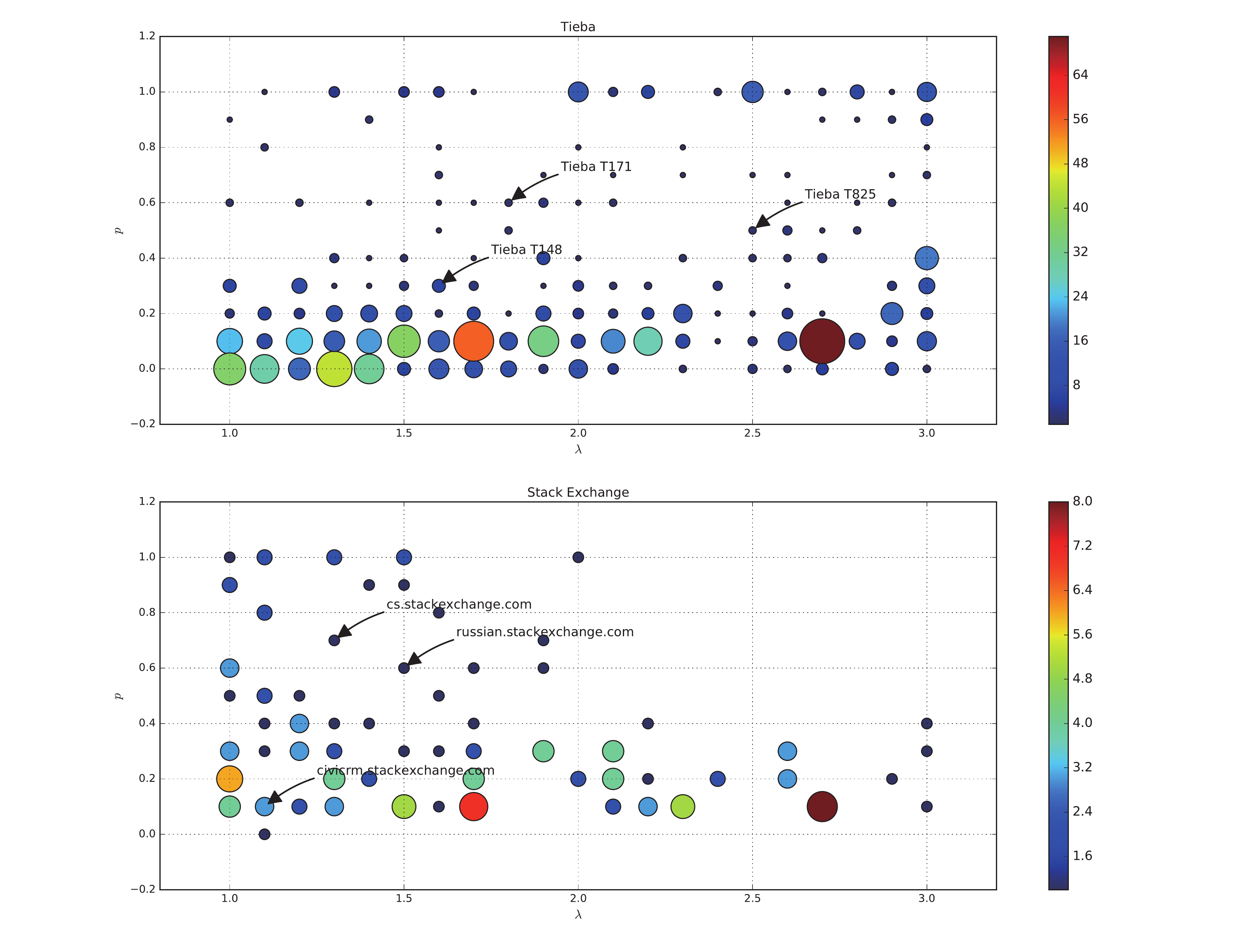}
	\caption{\textbf{Inferred individual parameters $p$ and $\lambda$ for all Baidu Tieba and Stack Exchange communities.} All the possible values of lambda and $p$ are separated into $10$ intervals, therefore, there are $100$ combinations of $p$ and $\lambda$ in total. We find the closest combination for each community. The size and the color both stand for the number of communities.}
	\label{fig:levy_stats}
\end{figure}

In figure \ref{fig:levy_stats} (a, b), we show all the inferred parameters for Baidu Tieba (a) and communities of Stack Exchange (b). We notice that all the Tieba’s can be roughly separated into two groups according to their parameters, and they have similar $p$ value ($0.1$) but different $\lambda$ values. We know that $\lambda$ indicates how dissimilar of the users’ interests for one transition. Thus, the users in Tieba with small $\lambda$ always have relatively wide interests. All the Tiebas’ have very small $p$ values meaning that the tendency for posting a new thread is much less than clicking. While, the communities in Stack Exchange almost concentrate in the area of $1.0<\lambda<2.0$ or $0<p<0.4$. That means the users in Stack Exchange always have wide interest and do not like to post questions. However, compared to Tieba, Stack Exchange communities always have larger $p$ values meaning that it is easier for asking a question compared to answering it than for posting a thread compared to clicking threads.

\section{Discussion}

In this paper we build an interactive L\'evy flight model to simulate the random walk behaviors of users in virtual interest space. We assume the users can interact indirectly via the digital resources. Two important parameters controlling the L\'evy flight’s behavior, i.e., how wide are users interests and the propensity that a user deliver a post determine the structures of the attention flow network. We compare the statistical properties of the attention flow network with empirical online communities from the perspective of scaling laws. Four different scaling laws characterize how the macroscopic quantities of activity, diversification of resources generated by users, and the diversification of interests transfer scale to the number of users. And the exponents characterize the relative growth speed. All the scaling behaviors and the range of exponents in simulation are in accordance with the empirical data. We then can infer the two important parameters $p$ and $\lambda$ if the exponents 

Therefore, the interest transition of users may be characterized by a simple random walk model on a 2-dimensional space spanned by the interests of users. The key that may explain the origins of the scaling laws that we have observed for the empirical communities is the indirect interactions between users. In our model, we assume that the users may stay in the system only if they can find the published digital resources which can feed their interests. This is the key to the indirect interaction and the super-linear scaling law of activity because when the number of users increases, the interactions between users also increase but in a faster rate. 

This work does not only provide theoretical understanding of online communities, but also implies potential applications. First, the scaling exponents can be treated as novel indicators to characterize the growth of communities. For example, the exponent $\alpha$ may indicate the level of interactive stickness of a community since it increases with the intensity of the interactions between users. The merits of adopting the exponents to quantify the communities include the stability of the exponent and the independency on the size of the community. Therefore, we can make a reasonable evaluation of a forum or a community when it is small.

Second, we can infer the parameters from the measured exponents. All these parameters describe the behaviors of users. Thus, our work makes it possible to infer the individual behavior only from their macroscopic performance of collective. And it is also possible that we can imply the macroscopic behavior if we know the individual parameters.

Third, we pave the path to connect the mobility between real and virtual worlds. Our model shows that human mobility in the virtual world may also follow the same statistical law as in the real world. And the interactions between people may play an important role.

Finally, drawbacks exist for the current model. First, we only provide an indirect evidence for the mobility in the virtual world. However, the space may not be 2-dimensional or even Euclidean. Second, the model simplifies the human behaviors in a large extent, this may not work if other factors need to be considered. Third, more empirical data should be collected to test our model.

\section*{Acknowledgments}
We gratefully acknowledge funding support from the National Natural Science Foundation of China (grant 61673070), the Fundamental Research Fund for the Central Universities(grant 310421103) and the Beijing Normal University Interdisciplinary Project.

\section*{References}
\bibliographystyle{iopart-num}
\bibliography{new_refs2}

\providecommand{\newblock}{}
\begin{thebibliography}{10}
\expandafter\ifx\csname url\endcsname\relax
  \def\url#1{{\tt #1}}\fi
\expandafter\ifx\csname urlprefix\endcsname\relax\def\urlprefix{URL }\fi
\providecommand{\eprint}[2][]{\url{#2}}

\bibitem{Grenfell2001Travelling}
Grenfell B~T, Bjørnstad O~N and Kappey J 2001 {\em Nature\/} {\bf 414} 716

\bibitem{Belik2011Natural}
Belik V, Geisel T and Brockmann D 2011 {\em Physical Review X\/} {\bf 1}
  3103--3106

\bibitem{Ratti2006Mobile}
Ratti C, Frenchman D, Pulselli R~M and Williams S 2006 {\em Environment and
  Planning B: Planning and Design\/} {\bf 33} 727--748

\bibitem{Lathia2012The}
Lathia N, Quercia D and Crowcroft J 2012 {\em Pervasive Computing\/} {\bf 7319}
  91--98

\bibitem{Wang2009Understanding}
Wang P, González M~C, Hidalgo C~A and Barabási A~L 2009 {\em Science\/} {\bf
  324} 1071--6

\bibitem{Ratti2010Redrawing}
Ratti C, Sobolevsky S, Calabrese F, Andris C, Reades J, Martino M, Claxton R
  and Strogatz S~H 2010 {\em Plos One\/} {\bf 5} e14248

\bibitem{Lathia2011Mining}
Lathia N and Capra L 2011 Mining mobility data to minimise travellers' spending
  on public transport {\em ACM SIGKDD International Conference on Knowledge
  Discovery and Data Mining\/} pp 1181--1189

\bibitem{Yuan2012Discovering}
Yuan J, Zheng Y and Xie X 2012 Discovering regions of different functions in a
  city using human mobility and pois {\em ACM SIGKDD International Conference
  on Knowledge Discovery and Data Mining\/} pp 186--194

\bibitem{Cacciapuoti2012Human}
Cacciapuoti A~S, Calabrese F, Caleffi M, Lorenzo G~D and Paura L 2012 {\em Ad
  Hoc Networks\/} {\bf 10} 1520--1531

\bibitem{Santi2013Taxi}
Santi P, Resta G, Szell M, Sobolevsky S, Strogatz S and Ratti C 2013 {\em
  Proceedings of the National Academy of Sciences\/}

\bibitem{Brockmann2006The}
Brockmann D, Hufnagel L and Geisel T 2006 {\em Nature\/} {\bf 439} 462--5

\bibitem{Gonz2009Understanding}
González M~C, Hidalgo C~A and Barabási A~L 2009 {\em Nature\/} {\bf 458}

\bibitem{Song2010Modelling}
Song C, Koren T, Wang P and Barabási A 2010 {\em Nature Physics\/} {\bf 6}
  818--823

\bibitem{Viswanathan1996L}
Viswanathan G~M, Afanasyev V, Buldyrev S~V, Murphy E~J, Prince P~A and Stanley
  H~E 1996 {\em Nature\/} {\bf 381} 413--415

\bibitem{Lomholt2008L}
Lomholt M~A, Tal K, Metzler R and Joseph K 2008 {\em Proceedings of the
  National Academy of Sciences of the United States of America\/} {\bf 105}
  11055

\bibitem{Raposo2009L}
Raposo E~P, Buldyrev S~V, Luz M~G~E~D, Viswanathan G~M and Stanley H~E 2009
  {\em Journal of Physics A Mathematical and Theoretical\/} {\bf 42} 434003

\bibitem{Rhee2011On}
Rhee I, Shin M, Hong S, Lee K, Kim S~J and Chong S 2011 {\em IEEE/ACM
  Transactions on Networking\/} {\bf 19} 630--643

\bibitem{Bartumeus2002Optimizing}
Bartumeus F, Catalan J, Fulco U~L, Lyra M~L and Viswanathan G~M 2002 {\em
  Physical Review Letters\/} {\bf 88} 097901

\bibitem{Lambiotte2008Geographical}
Lambiotte R, Blondel V~D, Kerchove C~D, Huens E, Prieur C, Smoreda Z and Dooren
  P~V 2008 {\em Physica A Statistical Mechanics and Its Applications\/} {\bf
  387} 5317--5325

\bibitem{Liang2009Avatar}
Liang H, Silva R~N, Ooi W~T and Motani M 2009 {\em Multimedia Tools and
  Applications\/} {\bf 45} 163--190

\bibitem{Wu2007Novelty}
Wu F and Huberman B~A 2007 {\em Proceedings of the National Academy of
  Sciences\/} {\bf 104} 17599

\bibitem{Wu2013The}
Wu L, Jiang Z and Min Z 2014 {\em Plos one\/} {\bf 9} e102646

\bibitem{Sasahara2013Quantifying}
Sasahara K, Hirata Y, Toyoda M, Kitsuregawa M and Aihara K 2013 {\em Plos
  One\/} {\bf 8} e61823

\bibitem{Kenett2014Discovering}
Kenett D~Y, Morstatter F, Stanley H~E and Liu H 2014 {\em Plos One\/} {\bf 9}
  e102001

\bibitem{Hawelka2014Geo}
Hawelka B, Sitko I, Beinat E, Sobolevsky S, Kazakopoulos P and Ratti C 2014
  {\em Cartography and Geographic Information Science\/} {\bf 41} 260

\bibitem{Li2014Game}
Li J, Theng Y~L and Foo S 2014 {\em Cyberpsychology Behavior and Social
  Networking\/} {\bf 17} 519--527

\bibitem{Etter2013Where}
Etter V, Kafsi M, Kazemi E, Grossglauser M and Thiran P 2013 {\em Pervasive and
  Mobile Computing\/} {\bf 9} 784--797

\bibitem{Watts2007Influentials}
Watts D~J and Dodds P~S 2007 {\em Journal of Consumer Research\/} {\bf 34}
  441--458

\bibitem{Smith2005Online}
Smith D, Menon S and Sivakumar K 2005 {\em Journal of Interactive Marketing\/}
  {\bf 19} 15--37

\bibitem{P2016The}
Pä‚Un C, Bratianu C, Pinzaru F and Zbuchea A 2016 {\em Management Dynamics
  in the Knowledge Economy\/} {\bf 4} 125--140

\bibitem{Liben2003The}
Liben-Nowell D and Kleinberg J 2003 The link prediction problem for social
  networks {\em Twelfth International Conference on Information and Knowledge
  Management\/} pp 556--559

\bibitem{Mislove2007Measurement}
Mislove A, Marcon M, Gummadi K~P, Druschel P and Bhattacharjee B 2007
  Measurement and analysis of online social networks {\em ACM SIGCOMM
  Conference on Internet Measurement 2007, San Diego, California, Usa,
  October\/} pp 29--42

\bibitem{Kumar2010Structure}
Kumar R, Novak J and Tomkins A 2010 Structure and evolution of online social
  networks {\em Link Mining: Models, Algorithms, and Applications\/} pp
  611--617

\bibitem{Cho2011Friendship}
Cho E, Myers S~A and Leskovec J 2011 Friendship and mobility:user movement in
  location-based social networks {\em ACM SIGKDD International Conference on
  Knowledge Discovery and Data Mining, San Diego, Ca, Usa, August\/} pp
  1082--1090

\bibitem{Shi2015A}
Shi P, Huang X, Wang J, Zhang J, Deng S and Wu Y 2015 {\em Plos One\/} {\bf 10}
  e0136243

\bibitem{Zeng2017Visualizing}
Zeng W, Fu C~W, Arisona S~M, Schubiger S, Burkhard R and Ma K~L 2017 {\em IEEE
  Transactions on Intelligent Transportation Systems\/} {\bf PP} 1--14

\bibitem{Frank2013Happiness}
Frank M~R, Mitchell L, Dodds P~S and Danforth C~M 2013 {\em Scientific
  Reports\/} {\bf 3} 2625

\bibitem{Yuan2015How}
Yuan T, Cheng J, Zhang X, Liu Q and Lu H 2015 {\em Knowledge-Based Systems\/}
  {\bf 88} 70--84

\bibitem{Pennacchiotti2011Investigating}
Pennacchiotti M and Gurumurthy S 2011 Investigating topic models for social
  media user recommendation {\em International Conference on World Wide Web,
  WWW 2011, Hyderabad, India, March 28 - April\/} pp 101--102

\bibitem{Cole2015User}
Cole M~J, Hendahewa C, Belkin N~J and Shah C 2015 {\em Acm Transactions on
  Information Systems\/} {\bf 33} 1

\bibitem{Middleton2004Ontological}
Middleton S~E, Shadbolt N~R and Roure D~C~D 2004 {\em Acm Transactions on
  Information Systems\/} {\bf 22} 54--88

\bibitem{Guo2015Flow}
Guo L, Lou X, Shi P, Wang J, Huang X and Zhang J 2015 {\em Physica A
  Statistical Mechanics and Its Applications\/} {\bf 437} 235--248

\end{thebibliography}

\end{document}